%% file: sample-sigconf.tex
\documentclass[sigconf]{acmart}
\usepackage{booktabs} 

\setcopyright{none}
\renewcommand\footnotetextcopyrightpermission[1]{}
\newif\ifshowcomments
\showcommentstrue

\ifshowcomments
\newcommand{\mynote}[2]{\textcolor{blue}{\fbox{\bfseries\sffamily\scriptsize#1}}
  \textcolor{blue}{{$/*$\textsf{\emph{#2}}$*/$}}}
\else
\newcommand{\mynote}[2]{}
\fi

\usepackage{textcomp}
\usepackage{xcolor}

\begin{document}
\title{Poster Abstract: LPWA-MAC - a Low Power Wide Area network MAC protocol for cyber-physical systems}

\author{Laksh Bhatia}
\affiliation{\institution{Imperial College London}}
\email{laksh.bhatia16@imperial.ac.uk}

\author{Ivana Tomi\'c}
\affiliation{\institution{Imperial College London}}
\email{i.tomic@imperial.ac.uk}

\author{Julie A. McCann}
\affiliation{\institution{Imperial College London}}
\email{j.mccann@imperial.ac.uk}


\begin{abstract}

Low-Power Wide-Area Networks (LPWANs) are being successfully used for the monitoring of large-scale systems that are delay-tolerant and which have low-bandwidth requirements.
The next step would be instrumenting these for the control of Cyber-Physical Systems (CPSs) distributed over large areas which require more bandwidth, bounded delays and higher reliability or at least more rigorous guarantees therein.
This paper presents LPWA-MAC, a novel Low Power Wide-Area network MAC protocol, that ensures bounded end-to-end delays, high channel utility and supports many of the different traffic patterns and data-rates typical of CPS.

\end{abstract}

%
%




\maketitle

\input{sections/Introduction}

\bibliographystyle{ACM-Reference-Format}
\bibliography{references}

\end{document}

%% file: sections/Introduction.tex
\section{Introduction}
Low Power Wide-Area Networks (LPWANs) \cite{Raza2017} are a new generation of wireless communication technology that have been designed to deliver long-range communications in the order of 10's of kms.
LPWAN technology offers a trade-off between power consumption, coverage, and data rates that make them ideal for monitoring in delay-tolerant applications that require small amounts of data to be transmitted periodically.
LPWANs are designed as low-power one-hop solutions which also makes them appealing as a communications strategy in the monitoring and control of Cyber-Physical Systems (CPSs) that are distributed over large areas.
A single-hop network topology is far more favourable for CPSs due to its relatively high predicable reliability and low energy cost compared to multi-hop networks.

\textbf{Challenges.} CPSs include physical processes which usually require higher data sampling rates when compared to applications for which LPWANs have been originally designed and this poses a challenge for the design of such protocols.
Also, differences lie in when data is sensed and communicated; there is an increasing tendency to design CPSs as event-based or event triggered systems where data is transmitted only when required.
This has the advantage of preventing unnecessary data transmissions and therefore reduces bandwidth and energy requirements, however, this also results in bursty transmission patterns with high-priority and time-critical data. LPWAN technologies have been designed for non-critical data communication that is periodic.

Among the existing LPWAN technologies, LoRa \cite{Raza2017} is regarded as one of the most promising from both, industry and academic communities.
It uses Chirp Spread Spectrum (CSS) modulation which makes the signals robust to interference and the entire LoRa system reliable.
LoRa uses LoRaWAN \cite{LoRaWAN2015} as its MAC protocol whose operating principles do not align with the requirements of a CPS.
LoRaWAN exploits a pure ALOHA channel access scheme which is unable to provide guarantees regarding maximum expected end-to-end delays and typically channel utilization is around 36\% \cite{Laya2016}.
Additionally, in most regions LoRaWAN operates under strict duty cycle regulations which prevents it from supporting bursty traffic patterns that are specific to CPSs and which results in long end-to-end delays and low reliability as shown in \cite{Tomic2018}.

\textbf{Contributions.} In this poster, we propose LPWA-MAC, an alternative to LoRaWAN that is able to handle the challenges of CPSs.
LPWA-MAC is a differentiated traffic protocol where resources are allocated on a request basis.
The traffic is differentiated into transmission channels and data slots based on traffic patterns and its timeliness.
By exploiting the advantage of multiple channels that are available in the 868MHz band and multiple data slots within each channel, LPWA-MAC achieves low end-to-end delays and high channel utility.
When compared to LoRaWAN, this implies a reduction of up to 84\% in end-to-end delays and an increase of up to 100\% in throughput.
Also, LPWA-MAC provides guarantees on maximum expected delay which is an absolute requirement for an on-time delivery of CPS's high-priority data. 
Further, while conforming to duty cycle regulations, it overcomes the problems of LoRaWAN by supporting high data-rates and bursty traffic patterns.

\section{LPWA-MAC Design}

The general design of LPWA-MAC protocol relies on several concepts, which are briefly presented in this section.

\textbf{Underlying Physical Layer.}
LPWA-MAC is designed as a link layer protocol for \emph{LoRa physical layer}; however, it is LPWA agnostic and it can be easily ported to the physical layers of many other LPWAN technologies. 

\textbf{Channel Access Scheme.}
LPWA-MAC exploits the notion of a \emph{traffic differentiation} in its channel access scheme.
The central authority (the gateway) differentiates traffic into several transmission channels and data slots based on traffic patterns and time criticality of data.
The resources (transmission channels and data slots) are allocated on a request basis.
The recent work in \cite{Zhang2017} has demonstrated superiority of queue-based channel access schemes, which are an example of the traffic differentiation, over traditional channel access schemes, such as CSMA and TDMA, in terms of end-to-end delay and throughput.
Additionally, in the case of LPWA-MAC the traffic differentiation ensures that the reliability of system is high and the number of data collisions that occur is zero.

\textbf{Resource Request Discipline.}
When a node wants to access the channel and send its data to the gateway, it needs to follow a \emph{request discipline}.
The node transmits its request for a data transmission to the gateway.
LPWA-MAC reserves a single channel for these uplink request transmissions and downlink request approval decisions.
If the request is granted, the gateway performs differentiation of node's traffic to a specific channel and data slot.
It also provides the \emph{time schedule} for node's data transmission.
Knowing its own schedule, the node can go to sleep until its time to transmit which minimizes the energy consumption.


\textbf{Distributed and Parallel Operation.}
LPWA-MAC uses \emph{distributed scheduling} which means that for a successful transmission the node does not need to know other node schedules.
This limits the communication overhead.
Additionally, \emph{multichannel data transmissions can run in parallel to the request procedures} which increases channel utilization and it further reduces end-to-end delays.
The addition of channel hopping means that if a node has bursty traffic it does not have to sleep for a regulated time in between the two consecutive messages to conform to the duty cycle regulation.
Rather it would just request new data slot and be migrated to a different channel where it would continue with its data transmissions.

\textbf{Handling Large Packet Sizes.}
To meet node's requirements for transmitting large packet sizes, LPWA-MAC supports \emph{multiple data slots allocation} to a single node.
This reduces the number of times a node needs to perform request discipline and its waiting time, which further reduces energy and end-to-end delays.



\section{Initial Results}

The current version of LPWA-MAC has been implemented in OMNeT++.
We compared its performance with LoRaWAN implementation that is based on Flora \cite{Flora2018} and that assumes message confirmation with up to $8$ retransmissions.
Both protocols have access to same communication resources (3 uplink and 1 downlink channel) while conforming to $1$\% duty cycle regulation.
The packet size is fixed to $40$Bytes and transmitted with spreading factor of $7$ which gives it a transmission time of $82$ms.
The data is generated according to a Poisson process where the network load varies from $2.5$ to $10$ packets per second. 
The number of nodes varies from $25$ to $200$. 

\begin{figure}[!t]
\begin{center}
\includegraphics*[width=0.88\linewidth]{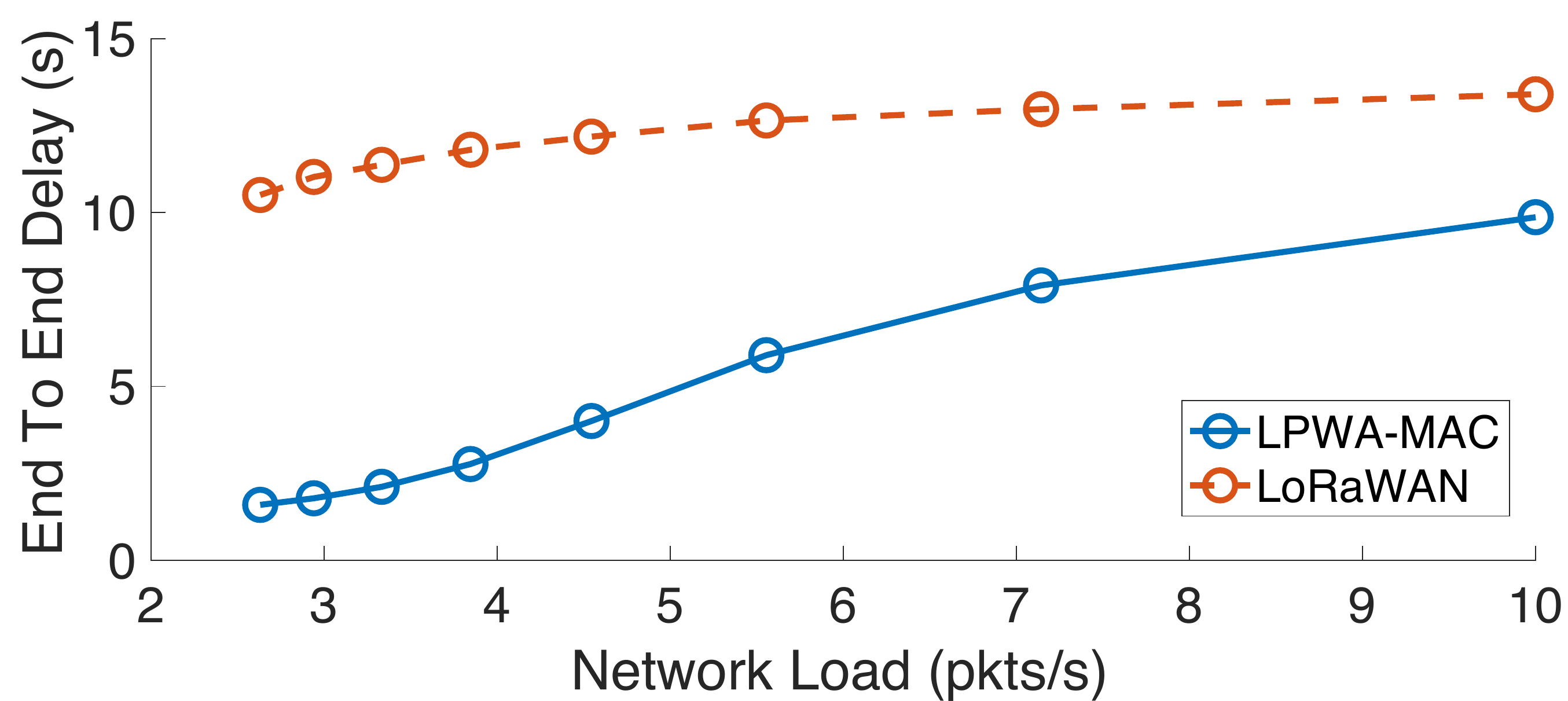}
\caption{\label{fig:E2Ed} Average end-to-end delay for $50$ nodes network and varied network loads.}
\end{center}
\end{figure}

\begin{figure}[!t]
\begin{center}
\includegraphics*[width=0.88\linewidth]{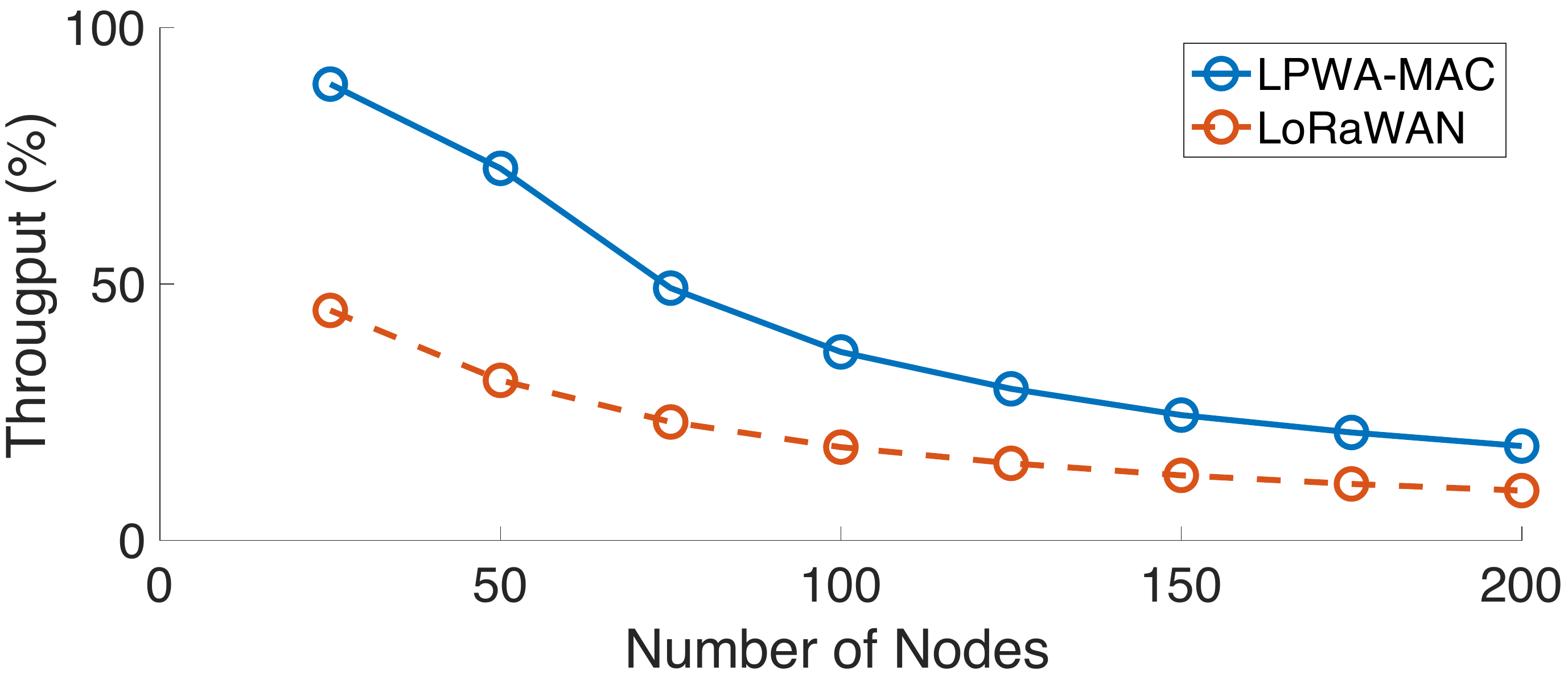}
\caption{\label{fig:thr} Average throughput for the network load of $4.5$ packets per second and varied network sizes.}
\end{center}
\end{figure}

We present two set of results.
Figure~\ref{fig:E2Ed} depicts the average end-to-end delay for $50$ nodes network where the network load is varied between $2.5$ and $10$ packets per seconds.
End-to-end delay is the time from sending a packet until its successful delivery at the gateway. The end-to-end delay for LoRaWAN does not change as the network size increases because on average
Figure~\ref{fig:thr} depicts the average throughput for varied network sizes ($25$-$200$ nodes) and fixed network load of $4.5$ packets/second.
The throughput is calculated as the amount of messages successfully sent by nodes over the total messages generated.
The results for LPWA-MAC show a reduction of up to 84\% in end-to-end delays when compared to LoRaWAN.
This shows a potential for supporting higher data rates for time-sensitive CPS scenarios.
There is also an increase of up to 100\% in throughput.
This has been achieved for non-periodical traffic patterns while conforming to duty cycle regulation.

\section{Conclusions \& Future work}

The initial set of results has demonstrated the potential to use LPWA-MAC in long-range CPSs where non-periodic traffic patterns with high-priority and time-critical data may occur.

We are currently working on testing the proposed protocol against varied traffic patterns and data-rates.
The design will be extended by the notion of priority slots which would allow us to satisfy requirements of both, delay-tolerant and delay-intolerant classes of applications, and allow them to share resources.
Additionally, LPWA-MAC will also take the advantage of multiple spreading factors that are available in the LoRa protocol.
This would further increase the channel utilization and the number of nodes that can be supported by a single LoRa gateway.

Finally, to validate LPWA-MAC and demonstrate its practicality, we will provide a real-world implementation using a water distribution system application that spans multi-kilometres.